# Spatially and polarization resolved plasmon mediated transmission through continuous metal films

Y. Jourlin,[1,*] S. Tonchev,[1] A.V. Tishchenko,[1]
C. Pedri,[1] C. Veillas,[1] O. Parriaux,[1] A. Last[2] and Y. Lacroute[3]

[1]*Université de Lyon, F-42023, Saint-Etienne, France,
CNRS, UMR5516, Laboratoire Hubert Curien, F-42000, Saint-Etienne, France,
Université de Saint-Etienne, Jean Monnet, F-42000, Saint-Etienne, France*
[2] *Forschungszentrum Karlsruhe, Germany*
[3]*LPUB UMR CNRS 5027, Université de Bourgogne, Dijon, France*
*[*yves.jourlin@univ-st-etienne.fr](mailto:yves.jourlin@univ-st-etienne.fr)*

**Abstract:** The experimental demonstration and characterization is made of the plasmon-mediated resonant transmission through an embedded undulated continuous thin metal film under normal incidence. 1D undulations are shown to enable a spatially resolved polarisation filtering whereas 2D undulations lead to spatially resolved, polarization independent transmission. Whereas the needed submicron microstructure lends itself in principle to CD-like low-cost mass replication by means of injection moulding and embossing, the present paper demonstrates the expected transmission effects on experimental models based on metal-coated photoresist gratings. The spectral and angular dependence in the neighbourhood of resonance are investigated and the question of the excess losses exhibited by surface plasmons is discussed.

______________________________________________________________________________________________

**References and links**
1. T. W. Ebbesen, H. J. Lezec, H. F. Ghaemi, T. Thio P. A. Wolff, "Extraordinary optical transmission through subwavelength hole arrays," Nature, **391**, 667-669 (1998).
2. F. I. Baida and D. Van Labeke, "Three-dimensional structures for enhanced transmission through a metallic film: Annular aperture arrays," Phys. Rev. B **67**, 155314 (2003).
3. N. Bonod, S. Enoch, Lifeng Li, Evgeny Popov, Michel Nevière, "Resonant optical transmission through thin metallic films with and without holes," Opt. Express **11**, 482-490 (2003).
4. I. Avrutsky, Y. Zhao, V. Kochergin, "Surface plasmon assisted resonant tunneling of light through periodically corrugated thin metal film," Opt. Lett. **25**(9), 595-597 (2000).
5. W. L. Barnes, S.C. Kitson, T.W. Preist, J.R. Sambles, "Photonic surfaces for surface-plasmon polaritons," J. Opt. Soc. Am. A **14**, 1654-1661 (1997).
6. S. Wedge and W. Barnes, "Surface plasmon-polariton mediated light emission through thin metal films," Opt. Express **12**, 3673-3685 (2004).
7. W.L. Barnes, A. Dereux and T.W. Ebbesen, "Surface plasmon subwavelength optics," Nature **424**, 824-830 (2003).
8. T. Inagaki, M. Motosuga, E. T. Arakawa and J. P. Goudonnet, "Coupled surface plasmons in periodically corrugated thin silver films," Phys. Rev. B **32**, 6238 - 6245 (1985).
9. Y. Jourlin, E. Gamet, S. Tonchev, A. V. Tishchenko, O. Parriaux, and A. Last, "Low loss polarizing beam splitter using the long range plasmon mode along a continuous metal film," Proc. SPIE **6187** (2006).
10. F. Pigeon, I.F. Salakhutdinov, A.V. Tishchenko , "Identity of long-range surface plasmons along asymmetric structures and their potential for refractometric sensors," J. Appl. Phys. **90**, 852-859 (1997).
11. D. Pietroy, A. V. Tishchenko, M. Flury, and O. Parriaux, "Bridging pole and coupled wave formalisms for grating waveguide resonance analysis and design synthesis," Opt. Express **15**, 9831-9842 (2007).
12. L. Li, J. Chandezon, G. Granet, and J. Plumey, "Rigorous and Efficient Grating-Analysis Method Made Easy for Optical Engineers," Appl. Opt. **38**, 304-313 (1999).

13. N. Lyndin. "MC Grating Software Development Company," http://www.mcgrating.com/  (April 2009).
14. I. F. Salakhutdinov, V. A. Sychugov, A.V.Tishchenko, B.A Usievich, O. Parriaux,  F. A. Pudonin, "Anomalous light reflection at the surface of a corrugated thin metal film," IEEE J. Quantum Electron. **34**, 1054-1060 (1998).
15. R. A. Innes and J. R. Sambles, "Optical characterisation of gold using surface plasmon-polaritons", 1987 J. Phys. F: Met. Phys. **17**, 277-287 (1987).
16. A. Degiron, P. Berini and R. Smith, "Guiding Ligth with Long Range Plasmon," Opt. Photon. News **19**, 7, 28-34 (2008).
17. P. Berini, R. Charbonneau, N. Lahoud, G. Mattiussi, "Characterization of long-range surface-plasmon-polariton waveguides," J.  Appl. Phys. **98** , 4, 043109.1-043109.12 (2005).
18. Benfeng Bai, Lifeng Li, and Lijiang Zeng, "Experimental verification of enhanced transmission through two-dimensionally corrugated metallic films without holes,", Opt. Letters **30**, 18, 2360-2362 (2005).

## 1. Introduction

Light transmission through metallic films can be caused by different mechanisms. The best known and most popular is that of propagation through periodic slits or holes perforating the metal film [1] where the transmitted spectrum is in the form of peaks resulting from resonances between in-plane and hole plasmon resonances [2]. It was shown however that light can also be transmitted through continuous metal films provided the latter is in the form of a deep and smooth undulation of adequate period [3]. It was also anticipated on the basis of a coupled mode vision, and demonstrated experimentally, that a shallow sinusoidal undulation can couple to the long range plasmon of the metal film and tunnel through the latter into the transmitted medium [4,5]. There is an abundant literature on the use of the plasmon tunnelling effect for extracting the light generated next to a metal electrode in LEDs and especially in OLEDs [6]. All these plasmon-mediated features rest on a solid theoretical basis built during the last two decades where the dispersion properties of grating coupled plasmons modes have been elucidated; most relevant references can be found in the review paper [7].

The present paper explores further the mechanism of plasmon tunnelling through shallow thin film undulations and does it in the practically highly interesting but very specific electromagnetic case of normal incidence where the coupled plasmons are standing waves. The coupling mechanism and related resonant transmission is applied to one-dimensional spatial filters exhibiting polarization and spectral filtering as well as simple polarization independent transmission through two-dimensional metal coated undulations. These filter functions can be implemented by means of high productivity injection moulding and embossing technologies. The application objective is to achieve the high resolution patterning and metal coating of a surface in the form of undulated and flat zones, the flat zones being essentially non-transmissive, the undulated zones being transparent in a given wavelength window and for a given polarization. The application objective is also to achieve a plasmon-mediated 2D structure which is polarization insensitive as an encodable information carrier. The problem of the large excess losses exhibited by a grating coupled plasmon is discussed on the basis of the obtained experimental results; although this problem was identified as early as in 1985 [8], it has not received a satisfactory explanation yet

## 2. The two plasmon modes along a thin metal film

A symmetrical structure will be considered here where the undulated zones and flat zones of a thin metal film are embedded in a polymer matrix composed of an injection moulded or embossed substrate and of a cover layer of thickness much larger than the wavelength. The substrate and cover have close to the same refractive index $n_c$. This is the basis structure of CDs where the pits are replaced by a sinusoidal corrugation; as it will be seen later, the needed corrugation depth is even smaller than the depth of a pit (50-70 nm instead of ca 120 nm). The metal of the sandwiched film has a complex permittivity $\varepsilon_m$.

A non-undulate metal slab propagates two TM plasmon modes as illustrated in Fig.1. The $TM_e$ mode with even magnetic and odd longitudinal electric fields and the $TM_o$ mode with odd magnetic and even longitudinal electric fields [9].

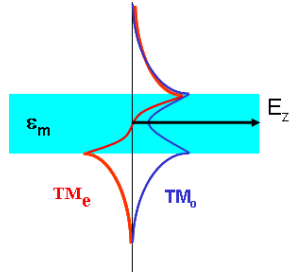

Fig.1: Longitudinal electric fields $E_z$ of the $TM_o$ (short range) and $TM_e$ (long range) modes propagating in an symmetrically embedded thin metal film

The dispersion equation satisfied by these two modes in a symmetrical structure is given by Eq. (1) for the $TM_e$ mode of even H field [10].

$$\tanh\left(\frac{k_m w}{2}\right) + \frac{k_c \varepsilon_m}{k_m n_c^2} = 0 \qquad (1)$$

For the $TM_o$ mode of odd H field, it is given by Eq. (2):

$$\tanh\left(\frac{k_m w}{2}\right) + \frac{k_m n_c^2}{k \varepsilon_m} = 0 \qquad (2)$$

where w is the metal film thickness, $\lambda$ the wavelength, $k_0 = \frac{2\pi}{\lambda}$, $k_m = k_0\sqrt{n_e^2 - \varepsilon_m}$ and $k_c = k_0\sqrt{n_e^2 - n_c^2}$. The effective index $n_e$ of the even mode is noted $n_{ee}$ and that of the odd mode $n_{eo}$.

It is well known that the effective index $n_{eo}$ of the $TM_o$ mode is the larger one which tends to infinity as the metal film thickness w decreases down to zero. The effective index $n_{ee}$ of the $TM_e$ mode is smaller and reaches $n_c$ as w tends to zero; there is no cutoff thickness for the $TM_e$ in a symmetrical structure. $n_{ee}$ tends to the cover index and can be approximated by Eq. (3) in cases of small metal thickness:

$$n_{ee} \cong n_c + \frac{1}{2n_c}\left(\frac{k_0(n_c^2 - \varepsilon_m) w n_c^2}{2\varepsilon_m}\right)^2 \qquad (3)$$

The $TM_e$ mode is called the long range plasmon whereas the $TM_o$ is the short range plasmon. The longe range plasmon mode can have especially low losses since the electron driving electric field $E_z$ is zero at the middle of the metal (whereas $H_y$ is even, $E_z$, proportional to the derivative of the latter, is odd, therefore has a zero crossing in the metal film), and the field tail extends deeply into the adjacent semi-infinite media.

Similarly to grating coupled dielectric slab waveguides, the reflection of a plane wave from an undulated metal film embedded in a homogenous medium is characterized in the neighbourhood of the mode coupling synchronism by a reflection maximum and a reflection minimum [11]. The reflection features from a corrugated metal film are best understood by first considering a lossless metal with zero imaginary part of the permittivity $\varepsilon_{mj} = 0$.

The synchronism coupling condition between the incident wave and a plasmon mode of effective index $n_p$ is $k_0 n_p = K_g \pm k_0 n_c \sin\theta_i$ where $k_0 = 2\pi/\lambda$ and $K_g = 2\pi/\Lambda$ is the grating constant, $n_c$ is the refractive index of the polymer matrix and $\theta_i$ is the incidence angle of a plane wave. The + and - signs stand for codirectional and contradirectional 1$^{st}$ order coupling respectively as illustrated in Fig. 2 where the grating $K_g$-vector is represented in the case of contra-directional synchronous coupling with the long range ($n_p = n_{ee}$) and the short range ($n_p = n_{eo}$) plasmons.

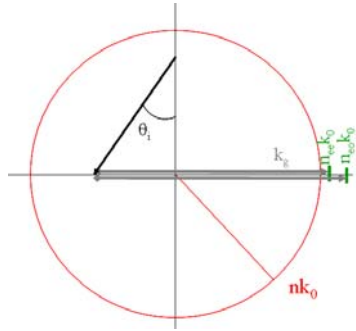

Fig. 2. k-vector diagram of the synchronism contra-directional coupling between the incident wave and the plasmon modes

Fig. 3 illustrates the modulus of the reflection coefficient at $\lambda = 890$ nm under $\theta_i = 30°$ incidence angle in a substrate of 1.6 index, from a lossless 30 nm thick gold film $\varepsilon_m = -31.7$ with a sinusoidal undulation of 70 nm amplitude upon a variation of the period $\Lambda$.

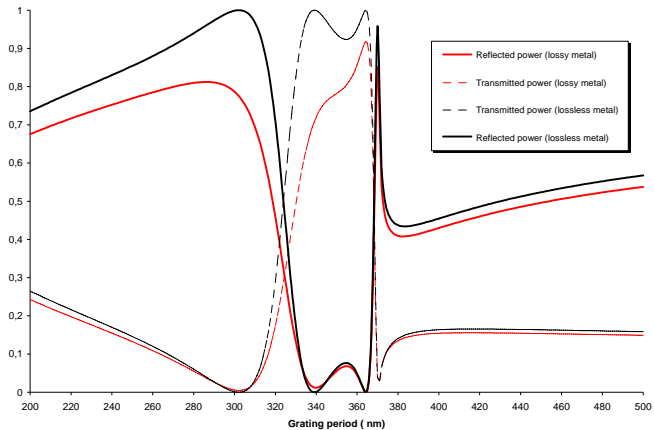

Fig. 3. Reflection and transmission spectra of an undulated metal film embedded in a homogenous medium versus the period $\Lambda$

The coupling in Fig. 3 is contradirectional in the -1$^{st}$ order to avoid the +1$^{st}$ diffraction order in the surrounding medium. For lossless metal (black curves), the reflection spectrum (solid black line) exhibits two 100% peaks and two zeros, the short period peak corresponding to the TM$_0$ short range plasmon. Unlike in a dielectric waveguides the minimum reflection here is always zero i.e. the transmission is 100% (dotted black line).

From Fig. 3, the effective index can be estimated by using the resonant contra-directional coupling condition for short- and long- range plasmons giving two grating periods (339 nm and 365 nm). $n_p = \lambda/\Lambda - n_c \sin\theta_i$ ($\theta_i$ being defined here in the incidence medium: $n_c = 1.6$). This gives $n_{ee} = 1.638$ and $n_{eo} = 1.821$.

Now introducing metal losses ($\varepsilon_m = -31.7 + 2j$) makes a significant change (red curves). The resonance positions are hardly changed but the reflection peak amplitudes decrease and the resonances widen as illustrated by the solid red line. The position of the zeros remains where it was. Remarkable is the fact that the long range plasmon is little affected by the introduction of the loss factor in $\varepsilon_m$: the reflection dip is still close to zero and the reflection peak is decreased to 90% (solid red line). The dotted lines represent the transmission in the presence of metal losses (red curve) and without losses (black curve). The transmission peak of the short range plasmon is wider and only reaches 70% which confirms that it is much more dependant on the imaginary part of the metal.

## 3. Experimental demonstration of the $0^{th}$ order transmission at normal incidence

Normal incidence is the most interesting configuration practically for a spatially resolved information carrier based on surface patterned, polarization dependent or independent transmission such as an encoder disk or a bar code. The above structure was submitted to an optimisation code based on the "C" method [12, 13] with the target of maximizing the transmission of the TM polarization as well as minimizing the TE transmission in the objective of optimizing the polarization contrast in undulated zones over a spectral width corresponding to that of a LED centered at 890 nm.

Under normal incidence, the synchronism condition writes simply $\Lambda = \lambda/n_e$. Because of the symmetry, the incident wave is coupled to plasmons propagating in both directions. This results in a plasmon standing wave. The result of the optimization with a lossy gold layer is shown in Fig. 4, where one recognizes the contribution of the two plasmon modes to the transmission versus the wavelength $\lambda$. The dotted black line is the transmission spectrum with lossless gold; it clearly shows two peaks of 100% resonant transmission and two transmission zeros which correspond to 100% resonant reflection. With a lossy metal film, a peak transmission of more than 90% is still expectable from the long range plasmon whereas it is only 60% from the more lossy $TM_0$ mode. The solid red curve represents the reflection spectrum of the TM polarization in the lossy case; summing up the TM reflection and transmission reveals that the metal losses are strong. This explains why abnormal reflection from an undulated film is so difficult to observe [14].

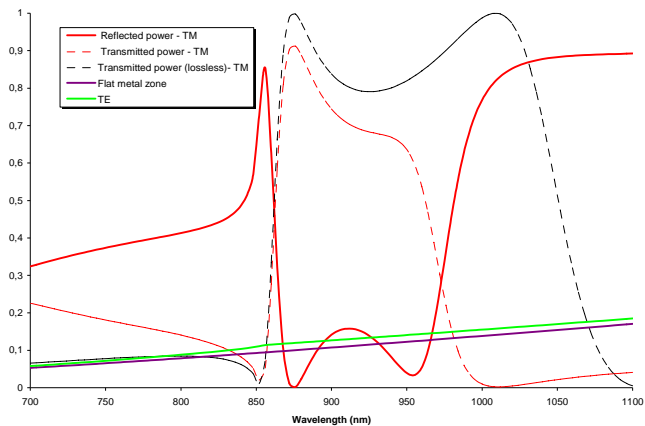

Fig. 4: TE and TM polarisation transmission trough a gold layer of 30 nm thickness under normal incidence versus wavelength. The period is 533 nm

The transmission of the TE polarization is kept under 15%. The TM and TE transmission of a flat metal film is very close to the latter as expectable since the corrugation is very shallow (70 nm for a period of 533 nm).

Although such structure can very easily be fabricated by means of a low cost CD technology, it is not so easy to prototype. An experimental simulation was made by first deposing a 1μm thick resist film on a glass plate, then performing a first uniform exposure by means of a mercury lamp followed by the exposure of an interferogram. The first exposure brings the resist ($n_c$ = 1.6) in the linear range implying that the interferogram prints as a sinusoidal undulation after development. A 30 nm gold layer was then deposited by evaporation without adhesion layer to prevent the large expectable losses in the plasmon propagation in the presence of a Cr or Ti adhesion layer. The same, but unexposed photoresist is then deposited as a 2μm thick cover layer. The grating only covers half of 50 nm diameter glass plates. Fig. 5(a) is the picture of a 2 inch diameter wafer after gold deposition on the sinusoidal corrugation and Fig. 4(b) is the AFM scan of the resist grating before gold deposition.

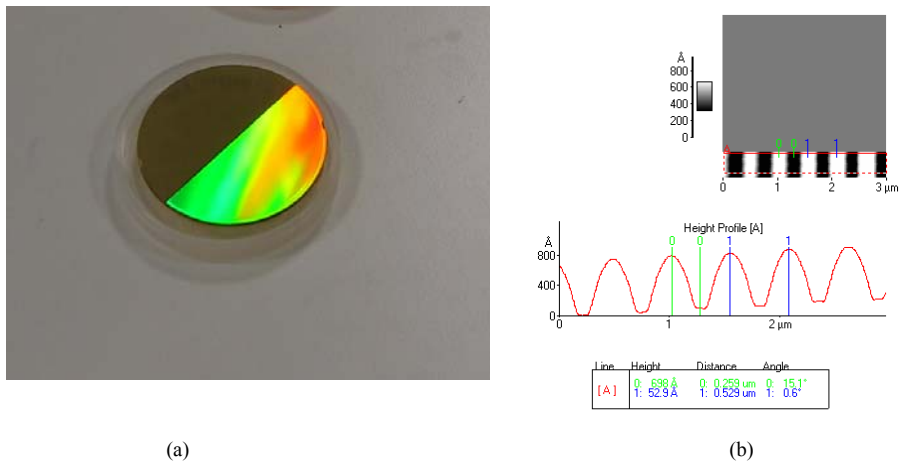

(a)            (b)

Fig. 5: Picture of a resist and gold coated glass wafer with the grating of 533 nm period and 70 nm depth covering half of the area (a) and AFM scan of the corrugated part (b).

The spectral transmission was measured by a Perkin Elmer spectrometer (Lambda 900) under normal incidence as illustrated in Fig. 6. The solid red line represents the TM transmission, the solid green line the TE transmission and the dotted line is for the TE and TM transmission trough a flat zone.

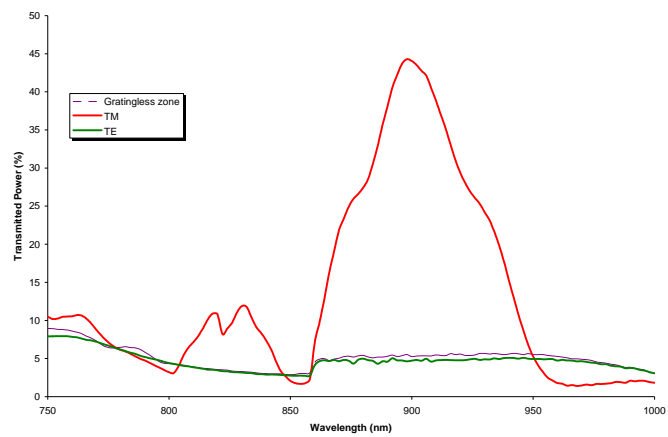

Fig. 6. Measured spectral transmission under normal incidence with an Aluminium layer

The comparison with the modelled structure shows that the peak TM transmission is smaller than 45 % whereas the TE transmission is close to the theoretical results. The resonance position is very close to 890 nm and one distinguishes the right shoulder of the $TM_0$ mode transmission. There are low amplitude TM transmission features at the short wavelength side of the peak. These correspond to TM guided modes propagating in the resist waveguiding multilayer of refractive index $n_c$ with sandwiched metal film surrounded by the glass substrate of lower index and by air at the cover side. Their effective index is smaller than $n_c$ therefore the resonance wavelength is smaller. Such features will disappear in future structures consisting of a metal film embedded in a fully homogenous medium.

In an attempt to increase the TM transmission peak, a gold layer was deposited by means of a slow gold evaporation (0.1 nm/s) in a vacuum below $5.10^{-8}$ mbar, at 18 degrees temperature, which are conditions known to lead to low loss plasmons. The result is represented by the curves of Fig. 7. The transmission is notably larger but remains well inferior to the theoretically expected 90%. We are facing here the important problem of excess losses which concerns most plasmonic devices. The scientific literature on this issue is relatively scarce [15, 16, 17]. Berini paper [17] reports that the experimental losses of his long-range plasmons are consistent with the metal bulk permittivity, provided that the metal thickness is larger than 22-23 nm. For smaller metal thicknesses, the losses become much higher than what is expected from bulk permittivity measurements. Based on a critical review of the literature on metal permittivities, Berini attributes this deviation to fabrication issues. The theoretically expected 90% transmission is based on the permittivity data available for bulk gold. In the scientific and industrial communities concerned with goldened gratings like femtosecond pulse compression gratings it is generally stated that the optical characteristics of professionally deposited gold layers are very close to those of bulk gold. If this would be also true for metal gratings involving plasmon excitation and propagation, one should find experimentally a resonant transmission much closer to 90%. Whether the excess losses are due to a surface plasmon wave being more sensitive to the metal film nanostructure than a free space wave experiencing non-resonant diffraction at the same metal grating or/and they are due to the imperfect wave front of a low spatial coherence light source is not known yet. It might be that the rather broad k-vector spectral width of a lossy plasmon of effective index close to the index of the embedding material induces a grazing free space wave leakage which increases in the presence of a surface undulation. The analysis of Ref. [8] in the case of a sinusoidal undulation three times shallower than in the present work and a period three times larger attributes the precisely measured excess losses to the very fact that the metal film is undulated. This calls for further theoretical work and fine characterization experiments and for a systematic exploration of metal film deposition technologies. This is beyond the scope of the present work which however confirms that the loss issue is a key issue which deserves further investigations.

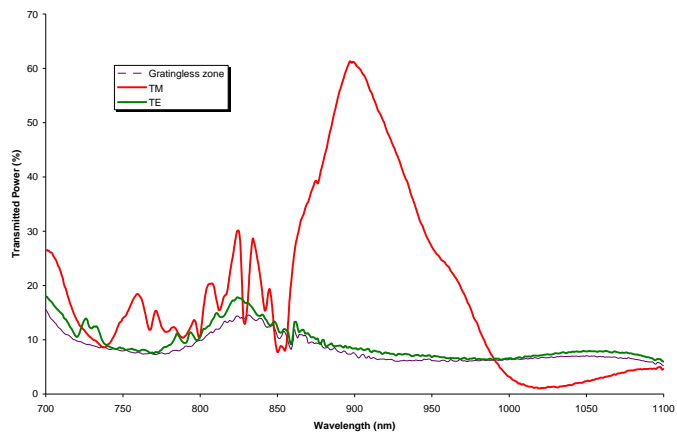

Fig. 7. Measured spectral transmission under normal incidence with a gold layer deposited by slow evaporation process.

## 4. The angular aperture

The information encoding devices which will use the effect of polarized or non-polarized resonant transmission will mostly use broadband sources of low or very low spatial coherence. Therefore the angular dependence of transmission around normal incidence is an important characteristics. If the angular aperture of the incident beam is larger than the angular aperture of the resonant effect, the resonant transmission peak amplitude will decrease.

A specific spectral measurement set up was built which uses a rotation stage with 0.5 degrees resolution installed in the spectrophotometer.

Fig. 8(a) represents the transmission spectrum of the TM polarization for different incidence angles. The transmission peak separates into two peaks at either side of the normal incidence central peak. Referring to the reciprocal space k-vector diagram of Fig. 2, this is explained by the fact that co-directional plasmon coupling takes place at a longer wavelength (smaller circle radii) whereas contra-directional coupling occurs at a shorter wavelength as represented more clearly in the modeling curves of Fig. 8 (b).

As in the short wavelength spectrum of Fig. 7, the spectra of Fig. 8 (a) exhibit a number of distinct transmission peaks and troughs. Again, these are the signature of fast wave modes guided in the resist metal sandwich and excited by the grating in the high frequency part of the spectrum. These experimental artifacts will disappear once the undulated metal film is embedded into a homogeneous polymer medium.

The mid- and long- wavelength part of the spectra is more interesting as it concerns the slow waves of the two plasmons excited contra- and codirectionally. Under oblique incidence there are two long range and two short range plasmons propagating in opposite directions. The curve of Fig. 8 (b) corresponding to normal incidence indicates that the effective index of the long and short range plasmons is 1.64 and 1.82 respectively. An angular offset of 1° leaves the excitation wavelengths of the contra- and co-directional long range plasmons close to each other: 854 and 884 nm respectively. The modelling curve resolves the two long range plasmons peaks and also the co-directionally coupled short range peak at 970 nm, but the experimental spectrum hardly resolves the two long range peaks. Interestingly, under 2° incidence, the experimental curve does resolve the two long range and the co-directional short range. Under 4° incidence, the spectral splitting of the contra- and co-directional plasmons is such that one only distinguishes the contra-directional long range peak at 810 nm and the co-directional short peak at 1000 nm wavelength; the co-directional long range and contra-directional short range interfere in the 900 nm wavelength range.

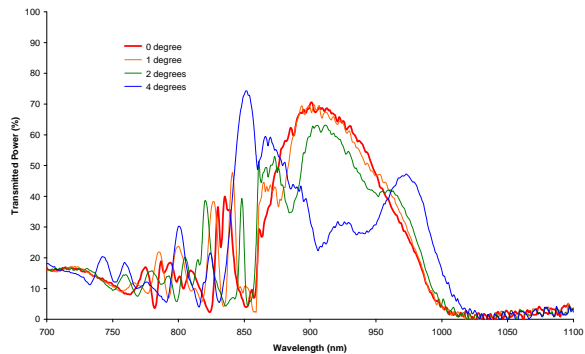

(a)

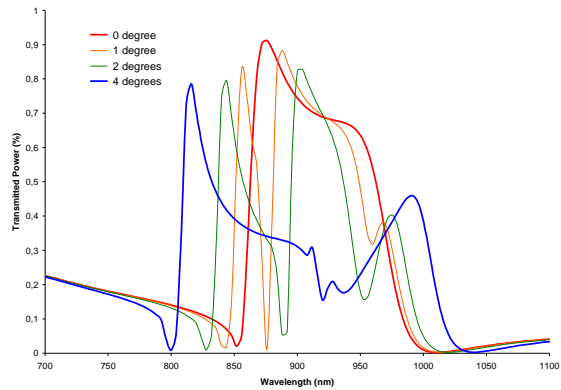

(b)

Fig. 8. Transmission power of the TM polarization versus the wavelength for different incidence angles. Experimental results (a) and modeling results (b). The period is 533 nm

Fig. 9 represents the angular dependence of the transmission through an undulated embedded metal film at the wavelength providing maximum transmission under normal incidence. The experimental curve results from a large number of spectra measured under different incidence angles as in Fig. 8 (a). The modelled angular spectrum using the bulk metal permittivity exhibits sharp transmission dips down to zero transmission as in Fig. 8 (b) in the spectral domain, i.e., a minimum transmission which is lower than the transmission through an non-undulated metal film. This situation of zero transmission corresponds to diffraction mid-way between the contra- and co-directional long range plasmon excitation and its clear that there is here a destructive interference in the transmission direction. Obviously, the experimental curve does not resolve such features. There are excess losses which broaden the angular spectrum and level off these fine artefacts. More efforts must be made to characterize, analyse and improve the angular spectrum; for the time being it can be said that resonant transmission effect has a full width of about 4-5 degrees.

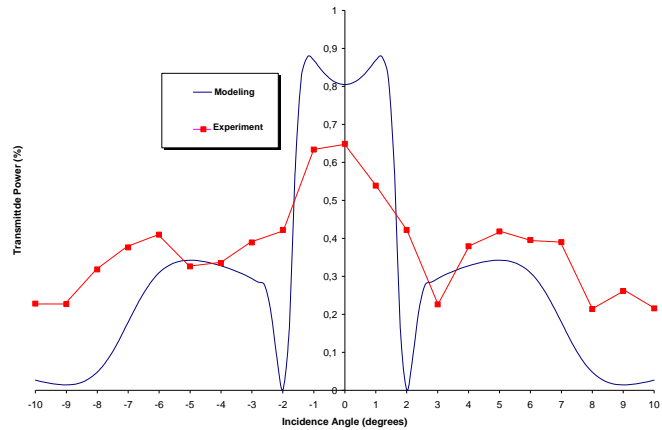

Fig. 9. Incidence angle dependence of the transmission at 890 nm wavelength. Grating period is 533 nm.

## 5. Polarization independent resonant transmission through a 2D metal grating

Resonant transmission can be made polarization independent by using a 2D sinusoidal grating of orthogonal lines [18]. Considering the data of the 1D grating as a starting point, a 2D grating code based on the Rayleigh Fourier approximation searches for the resonance condition and optimises the grating depth maximizing the ratio between TM and TE transmissions.

Depending on the resist exposure conditions, the 2D grating can be a mix of two distinct profiles: a 2D corrugation expressed as the sum of two orthogonal cosinus functions of the same argument as the 1D grating or as a product of them. The latter corresponds to a structure where the dominant diffraction orders are contained in two planes parallel to the diagonals of the resulting chessboard with an effective period of $\Lambda/\sqrt{2}$. The former corresponds to a structure where the dominant orders are along the x and y directions with period $\Lambda$.

The grating printing technique used here was a first uniform UV flood for placing the resist in its linear regime, followed by two interferogram projections at 90 degrees. The resulting surface profile is represented by the 2D AFM scan of Fig. 10 exhibiting a very good uniformity from where it is however difficult to determine the type of x,y undulation function. It should be of the type of a sum of cosine functions; one can however distinguish a flattening of the tops; this could give rise to diffraction orders along the diagonals.

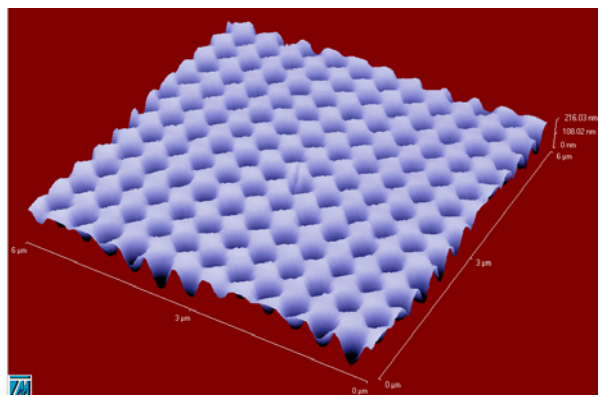

Fig. 10. AFM scan of a 2D resist grating printed by two successive orthogonal 1D exposure after uniform UV flood. The period of each 1D interferogram is 533 nm.

An non-resonant optical diffraction test was made after resist development in a beam at 442 nm wavelength. At 442 nm wavelength the + and − 1st orders generated by the spatial period 533 nm do propagate in the air whereas those generated by a period of 533/√2 = 377 nm are evanescent in the air and only propagate in the substrate of the 2D grating. Figure 11 shows the 4 bright ± 1st transmission orders corresponding to the period Λ = 533 nm of the same 2D grating. The evaluation of the diffraction efficiency of the + and − 1st orders along the chess board diagonals corresponding to the 377 nm period was made by resorting the oblique incidence. These orders can be seen, but their intensity is about one order of magnitude lower. This means that the 2D grating is like a sum of two cosinus of dominant period Λ = 533 nm with however some flattening of the profile tops generating weak orders along the diagonals.

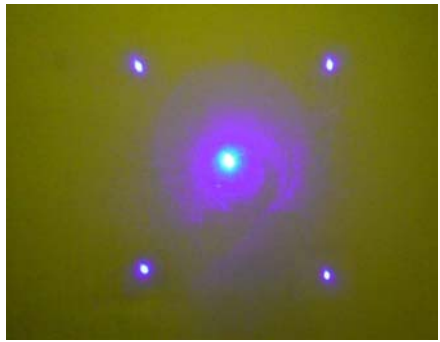

Fig 11. Picture of the diffracted orders generated in air by the 2D grating. The lines of the orthogonal intreferograms are along the diagonals of the picture.

After gold and aluminum coating of 30 nm and 20 nm thickness respectively and resist overlay spin-coating, the transmission spectra demonstrate as shown in Fig. 12(a) and 12(b) that the transmission peaks are still there at the same spectral position with a non-polarized beam in the 2D grating zone whereas it is between 5% to 10% transmission for non-undulated zones. The experimental results show that 0th order transmission in a symmetric structure was achieved with almost 30% for an Aluminum film and 60% for a gold structure which is comparable to the experimental transmission obtained in the 1D case for the TM polarization. In the present R&D work silver has not been considered as a possible layer material although silver is a low loss optical metal which exhibits dispersion properties permitting an easy comparison with between theory and experiment. A silver layer in a real life environment would quickly loose its optical properties once exposed to permeating traces of sulphur and chlorine even though the ultra-thin film would be embedded in a polymer.

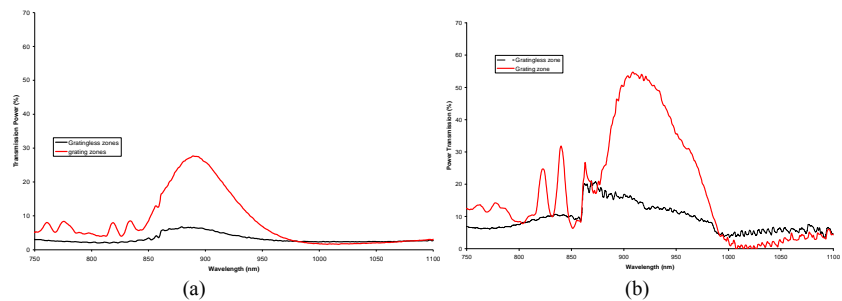

(a)     (b)

Fig. 12. Measured spectral transmission under normal incidence on 2D gratings. Aluminum grating (a) – Gold grating (b).

The described effect of resonant transmission through a patterned surface can be used in many applications as an information carrier technology. Beyond the possible use of spectral multiplexing, the main interest is the fabrication cost: such subwavelength patterned element can be fabricated by means of CD-technology or any low cost replication technology. The example of a functional structure is given in Fig. 13 which suggests that the binary transmission/reflection function of a patterned chromium-coated glass disk can be advantageously performed by an uniformly metal-coated polymer disk having replicated submicron undulated windows and flat reflecting zones. Fig. 13(a) is the picture of a conventional optical disk and Fig. 13(b) is the AFM scan of the substrate with a 2D grating and a flat zone corresponding to the transmissive and reflective zones respectively. It is interesting to note that an aluminum film is a good candidate for resonant transmission applications despite the smaller achievable transmission: the contrast between the transmission through undulated and flat zones is not significantly smaller than in the case of a gold film. The interest of aluminum is reinforced by the fact that its adhesion to polymers is notably stronger than for gold.

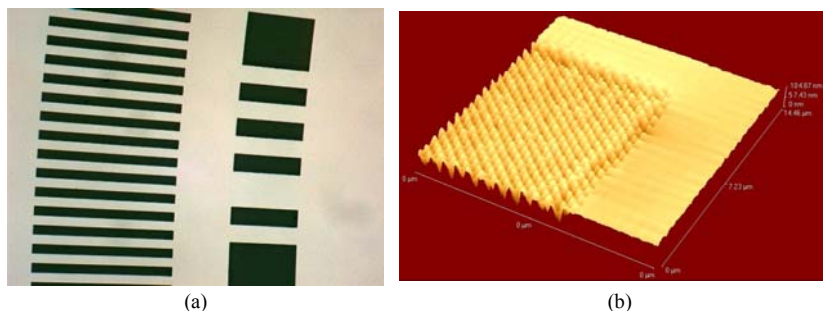

Fig. 13. Optical microscope picture of an optical disk (a) and AFM scan of the 2D grating and flat zones (b)

## 6. Conclusion

Surface plasmon-mediated resonant transmission through undulated continuous metal films under normal incidence was shown to enable a spatially resolved as well as polarization selective digital encoding of spatial resolution of a few micrometers. Although the demonstration and characterization of the resonance effect was made on an experimental model using photoresist prototyping, the simplicity of the structure lends itself to low-cost CD-like replication technologies. This has the somewhat paradoxical consequence that spatially resolved functions as simple as submillimeter resolution reflection/transmission which are usually implemented by membrane hole piercing, ink patterning, laser darkening, etc, can now be performed at a lower manufacturing cost by means of replicated submicron, electromagnetically resonant features. The resonant character of the transmission effect permits to possibly multiplex several wavelength channels at the same surface. The polarization selectivity of spatially resolved 1D undulations permits to add one more level of functional complexity and to give rise to polarizing elements which do not have their classical counterpart like azimuthal and radial linear polarizers.

This potential can be exploited provided a substantial research effort is devoted to the loss issue in plasmonics. The present paper has brought the evidence that plasmon-based metal optics faces excess losses whose origin still has to be studied before technological solutions can be engineered.

### Acknowledgements

This work was made in part in the framework of the Network of Excellence of the EC on micro-optics NEMO.